\documentclass[onecolumn,secnumarabic,nobalancelastpage,nofootinbibt]{revtex4-1}

\usepackage{lineno}
\usepackage{graphics}      
\usepackage{graphicx}      
\usepackage{longtable}     
\usepackage{url}           
\usepackage{bm,color}            
\usepackage{amssymb, amsmath, enumerate, theorem, epsfig,subfigure,setspace}
\usepackage{amsfonts}
\usepackage{mathtools}
\usepackage{listings}
\usepackage{float}
\begin{document}

\title{Benchmarking the optimization optical machines with the planted solutions}
\author{Nikita Stroev$^{1}$}
\email[correspondence address: ]{nikita.stroev@weizmann.ac.il}
\author{Natalia G. Berloff$^{2}$}
\author{Nir Davidson$^1$}
\affiliation{$^1$Department of Physics of Complex Systems, Weizmann Institute of Science, Rehovot 76100, Israel}
\affiliation{$^2$Department of Applied Mathematics and Theoretical Physics, University of Cambridge, Cambridge CB3 0WA, United Kingdom}

\begin{abstract}

We introduce universal, easy-to-reproduce generative models for the QUBO instances to differentiate the performance of the hardware/solvers effectively. Our benchmark process extends the well-known Hebb's rule of associative memory with the asymmetric pattern weights. 
We provide a comprehensive overview of calculations conducted across various scales and using different classes of dynamical equations. Our aim is to analyze their results, including factors such as the probability of encountering the ground state, planted state, spurious state, or states falling outside the predetermined energy range. Moreover, the generated problems show additional properties, such as the easy-hard-easy complexity transition and complicated cluster structures of planted solutions. Our method establishes a prospective platform to potentially address other questions related to the fundamental principles behind device physics and algorithms for novel computing machines.

\end{abstract}

\maketitle

\section{Introduction}
\label{Introduction}



Combinatorial optimization (CO) has always been an important subfield of applied mathematics due to its close connection with many practical problems. The quadratic unconstrained binary optimization (QUBO) is a specific type of CO problem with a large accumulated number of methods for solving it, such as simulated annealing, genetic algorithms or various heuristics, see \cite{dunning2018works} and \cite{kochenberger2014unconstrained} for an overview of the various problem types and solution techniques. QUBO has attracted a lot of attention due to its simplicity, its applicability to a broad range of NP-hard CO problems together with machine learning (ML) applications \cite{lucas2014ising,mucke2019learning} and most importantly, QUBO correspondence to Ising Hamiltonians \cite{hertz2018introduction}. This equivalence is twofold, leading to the conversion of the applied problems into combinatorial optimization and the ability of various physical platforms to solve the QUBO problems without additional sophisticated tuning of the system. That is why QUBO also appeared among the first tasks to be attacked by the practical realizations of unconventional hardware.

The latest progress in developing various unconventional computing platforms allows one to obtain efficient ways to address optimization problems by leveraging the classical physical mechanisms, significantly lowering computational resource requirements \cite{minzioni2019roadmap,woods2009optical,wu2021artificial,stroev2023analog}, which resulted in the optical and other physics-based hardware emergence. We will focus only on the systems that admit quasi-classical description, although the quantum effects can foster their functionality. Among these platforms are the exciton-polariton condensates \cite{BerloffNatMat2017,NJP_Kalinin2018}, coupled laser systems \cite{lasers3,nirprl2017,nir2018}, the coherent Ising machine (CIM) consisting of optical parametric oscillators (OPOs) \cite{mcmahon2016fully,CIM3,CIM4,CIM8} and many others \cite{mourgias2023analog,stroev2023analog}. These platforms operate using different physical mechanisms and have their own strengths and weaknesses. However, their classical description can admit similar mathematical descriptions with the corresponding algorithmic realizations \cite{NJP_Kalinin2018,stroev2021discrete,chermoshentsev2021polynomial,tiunov2019annealing}.

The abundance of various platforms naturally leads us to the problem of developing benchmarks. One can find it hard to compare the performance of many existing devices and algorithms due to a  lack of uniformity among them. The heuristic nature of the results, the increasing number of specific algorithms and corresponding modifications, the task-specific benchmarks and problem instances that can be potentially exploited for the biased results' presentation are all contributing to the difficulty of the cross-comparison. This work's primary motivation is to develop an effective benchmarking process for the special-purpose hardware and algorithms to evaluate and differentiate their performance effectively. Introducing the universal generating models should reduce the overwhelming variety of instance parameters. In addition, it is possible to address many other model-specific questions, such as a precise characterisation of the individual physical trajectories and their deviation from particular practical realizations in the hardware or addressing the role of specific quantum effects.

One of the benchmarks for the initial testing of the CIM was the specific Möbius ladder graph instance \cite{mcmahon2016fully}. However, it appeared that the minimisation of Ising Hamiltonian on such graphs does not pose serious difficulty, and many optical optimization machines show a good performance using such instances. This problem can be made harder by introducing the rewiring procedure for the connectivity graphs to increase the complexity of the problem and the specific simplicity criteria to measure this complexity \cite{kalinin2022computational}. The statistical approach is another way to measure the complexity of the computational problems \cite{zdeborova2008statistical,zdeborova2016statistical}. The complexity can be elucidated in the vicinity of the easy-hard-easy complexity transition in the SAT tasks \cite{gent1994sat} or in the generalization in neural networks (NNs) \cite{aubin2018committee,gerace2020generalisation}. This statistical approach draws the correspondence between models of phase transitions in physics and complexity transitions in computational problems. Wishart planted ensemble \cite{hamze2020wishart} and 3D tiling problem \cite{hamze2018near} were proposed as the problem instances with a tunable hardness to address the benchmarking issues. The statistical approach also has many results concerning inference \cite{zdeborova2016statistical}, ML-related tasks \cite{krzakala2021statistical,abbaras2020rademacher,barbier2019optimal} and compressed sensing \cite{krzakala2012probabilistic,donoho2009message}, that later was used to reevaluate the performance of the CIM \cite{aonishi2022l0}. 
The research on different optimization problems over random structures beyond QUBO condensed in the general criteria of statistical hardness. A new approach for algorithmic intractability is called the Overlap Gap Property and is based on the topological disconnectivity property of the set of pairwise distances of near-optimal solutions \cite{gamarnik2021overlap}. It emerges in many models with the prior random structures, coincides with the conventional hardness phase transition and is related to the stability of the algorithms. This property can be applied to the description of the hardware operating principles. See also the review article \cite{gamarnik2022disordered}, which highlights connections between the physics of disordered systems, phase transitions in inference problems, and computational hardness.

However, many of the suggested benchmarking instances have specific drawbacks and cannot characterise the physical systems' evolution. Some of them are specifically tailored to particular hardware to highlight its strengths (e.g. Möbius ladder instances) or inherently possess statistical properties that make it hard to analyze (e.g. Wishart planted ensemble or 3D tiling), i.e. to characterize the solution space properties (e.g. the unified framework via generic tensor networks in \cite{liu2022computing}). We aim to construct instances not only with controllable hardness, but also with controllable distances between clusters of low-energy solutions and the energy difference between them. Our construction uses the methodology based on the well-known associative memory model \cite{hebb1949organization,hopfield1982neural} with additional modifications for the optimization context. It has similar advantages and also eliminates many drawbacks of the previous models. By introducing asymmetry among the planted memory patterns, we eliminate the degeneracy between multiple ground states. This not only allows for more expressive representations of results in various optimization contexts but also enables easy comparison using the corresponding distributions. Moreover, it is possible to study the solution space properties and even go beyond, i.e. to study individual dynamical trajectories and transformations of the phase space of possible solutions. 

This paper is organized as follows. Section \ref{Toy model for the complex landscape and dynamical equations} is devoted to the mathematical description of the QUBO problem, the parametrization of our model and dynamical equations used to solve QUBO instances. Section \ref{Small scale classification} contains the results concerning the small-scale problems and their complete description. Section \ref{Numerical results for large N} extends the proposed approach to medium and large-scale problems with the corresponding numerical results and the additional observed features. Discussion and conclusions are given in Section \ref{Discussion and conclusions}. Finally, Supplementary Section provides additional small-scale numerical experiments and additional information on the problem details and hardware adjustments.

\section{Toy model for the complex landscape and dynamical equations}
\label{Toy model for the complex landscape and dynamical equations}

QUBO is a CO problem with a wide range of applications in many fields, such as finance and economics, physical sciences, and its original domain - computer science \cite{kochenberger2014unconstrained}. The majority of the works on QUBO are devoted to the development of the different heuristics \cite{dunning2018works} and implementation of QUBO on particular physical devices \cite{stroev2023analog}. 


One of the common formulations of the QUBO problem has the following form:

\begin{linenomath*}
\begin{equation}
\min _{\mathbf{\tilde{x}} \in\{0,1\}^N} E = \sum_{i,j \geq i} {P}_{ij} \tilde{x}_{i} \tilde{x}_{j},
\label{qubo}
\end{equation}
\end{linenomath*}

where one looks for a vector of binary variables with $N$ components $\tilde{x_i}^{*}$, which are coupled through the upper triangular matrix $P_{ij}$, that minimizes the given quadratic form. Through this paper we are interested in both identifying the precise minimizer for the given problem and tackling challenging instances by seeking approximate solutions that closely align with the values of the cost function at the minimizer.

QUBO is closely related and computationally equivalent to the Ising model \cite{lucas2014ising}, for which the task of finding the ground state of the Hamiltonian is written in the following manner:

\begin{linenomath*}
\begin{equation}
\min _{\mathbf{x} \in\{-1,+1\}^N} E = -\sum_{i,j>i} {J}_{ij} x_{i} x_{j},
\label{ising}
\end{equation}
\end{linenomath*}

where we omitted the term with the external magnetic field $ -\sum_j h_j x_j$ that we will not use in our context. Here the $N$ binary variables can take values from $\{-1,+1\}$ and are coupled through the coupling matrix $J_{ij}$. By applying the identity $x_i = 2 \tilde{x_i} -1$, one can get the correspondence between these two equivalent problems through the following coefficients relations: $P_{ij} = -4J_{ij}, i \neq j$; $P_{ij} = 2\sum_{k \neq j}J_{kj}-2h_{j}, i=j$ and constant for the QUBO formulation which equals $-\sum_{i>j} J_{ij} + \sum_{j} h_j$ and does not change the position of the minimizer. By referring to the QUBO in the majority of situations in this paper, we will mean its Ising formulation, as given by Eq.~(\ref{ising}), alongside the variables, referred to at times as spin-variables or simply spins.

QUBO is one of the computational tasks that can be realized by analogue computing devices (although the effectiveness of these platforms may be limited by specific hardware constraints, i.e. not every QUBO problem can be straightforwardly transferred to analog devices of such kind) and is quite popular due to its simple formulation, connections to spin glass models and Ising model in particular and well-established efficient connections with many combinatorial optimization problems such as maximum cut, graph colouring or the partition problem \cite{lucas2014ising}. The QUBO applicability for various machine learning embeddings \cite{mucke2019learning} is another factor in its popularity.

QUBO is an NP-hard problem \cite{kochenberger2014unconstrained} with the number of possible solutions growing as $2^{N}$. Generally, there are no straightforward tools to reduce the overall growth of complexity in the QUBO problem.
To tackle QUBO, different heuristics were invented and are still an active area of research \cite{kochenberger2014unconstrained,schuetz2022combinatorial,mohseni2021nonequilibrium}.

There are many well-known mathematical extensions of the QUBO problem, such as HOBO (Higher-Order polynomial Binary optimization) \cite{stroev2021discrete,chermoshentsev2021polynomial}, QUBO with the additional magnetic field term \cite{kadanoff2000statistical}, or additional constraints on the space of potential solutions (that dramatically change the utility of various methods and heuristics) and complex QUBO analogues such as minimization of the XY model (easily accessible by many optical systems) \cite{BerloffNatMat2017,nir2018}. Although every problem can be converted into QUBO by straightforward discretization (to be more precise to HOBO and only then to QUBO with the additional auxilary variable, although with the change of the solution space), one has to pay a significant price with the overhead produced by the number of spin variables that can reach large numbers. 
One more benefit of the QUBO is its ability to be identified as a component within an alternative problem context or structure, which can make the special-purpose hardware helpful in this context, for example \cite{wilson2021machine}.

Characterizing random energy landscapes is a longstanding topic in many fields \cite{ros2022high,wolynes2001landscapes,austin2012free,krugman1994complex,goldstein1969viscous}. However, we are not only interested in the correlated problem instance (in comparison with the majority of the uncorrelated ones) but also in tracking the individual behaviour of various outcomes and their precise characterization.
Our model is based on Hebb's rule \cite{hebb2005organization}, which originally was introduced to describe the influence of neuronal activities on the connection between neurons, i.e. to characterize the synaptic plasticity. 
One of its popular formulations defines the coupling coefficients in Eq.~(\ref{ising}) as:

\begin{linenomath*}
\begin{equation}
J_{i j}= \sum_{m=1}^K \xi_i^m \xi_j^m,
\label{hebb}
\end{equation}
\end{linenomath*}

for $K$ planted binary patterns $\xi_i^{m}$ ($m=1..K$) with $i, j \in\{1, \ldots, N\}$. The most common use of this rule can be found in the associative memory models \cite{hopfield1982neural,little1974existence}.

Alternatively, Hebb's rule is a way to train weights in a NN without the sophisticated training procedure (e.g. back-propagation, the gradient estimation method widely used in artificial NNs) has originally appeared in a different context \cite{hebb1949organization}. The capacity (the amount of the possible stored patterns) of the Hopfield NN (Ising model of a NN or simple recurrent shallow NN) with the weights defined by Eq.~(\ref{hebb}) is given by $C \cong \frac{N}{2 \log _2 N}$ \cite{hertz2018introduction}, which depends only on the number of variables $N$ and is a well-known result in statistical learning theory. From the dynamical point of view, adding $K+1$ patterns is like adding another point with its basin of attraction (i.e. attractor) to the energy landscape together with its mirror vector ($\xi_i^{\mu \operatorname{mir}}=-\xi_i^{\mu}$) that appears to the multiplication symmetry of Eq.~(\ref{hebb}). It produces mixed patterns in the form $\xi_i^{\operatorname{mix}}=\operatorname{sgn}\left(\pm \xi_i^{\mu_1} \pm \xi_i^{\mu_2} \pm \xi_i^{\mu_3}\right)$ in case of $3$ planted patterns  (and other combinatorial variations of an odd number of patterns for $K>3$) \cite{hertz2018introduction}. There are many ways of using Hebb's rule for practical purposes; e.g. for the dense associative memory \cite{krotov2016dense} and in modern Hopfield networks \cite{ramsauer2020hopfield}.

One can modify the capacity of the Hopfield NN by applying the so-called pseudoinverse rule:

\begin{linenomath*}
\begin{equation}
J_{i j}= \sum_{\mu \nu} \xi_i^\mu\left(Q^{-1}\right)_{\mu \nu} \xi_j^\nu,
\label{pseudoinverse}
\end{equation}
\end{linenomath*}

where matrix $Q_{\mu \nu}=\frac{1}{N} \sum_i \xi_i^\mu \xi_i^\nu$ represents the overlaps between patterns $\nu$ and $\mu$ with $i, j \in\{1, \ldots, N\}$. Equipped with such a rule, one can achieve the capacity of a NN close to $C_Q \cong N$. Summing all the $N$ orthogonal patterns will lead to the $J_{ij}$ being zero, known as the saturation effect. In the case of the interactions of higher-order $n$, the capacity scales as $N^{n-1}$ \cite{krotov2016dense}. Here, we outline the process for constructing orthogonal patterns in our setup and explain the orthogonality property arising from this method. First, one has to choose $N = 2^d$, where $d$ is an integer. Then to generate a $2 \times 2$ binary matrix $((1,1),(1,-1))$ and create two copies of this matrix. For the first copy, one has to replace each element by its multiplication with the row $(1,1)$. For the second copy, one has to replace each element by its multiplication with the row $(1,-1)$. 
Recursively repeat the process ($d-2$ more times) with the obtained matrix until reaching the desired size $N$. This pattern continues for larger N, maintaining the orthogonality property throughout the construction process.
Additionally, one can notice that the rows of the generated matrix mirror those of the Hadamard matrix $H_{N}$, but in a different order. Essentially, there exist a permutation matrix $O_{ij}$, such that the expression $O_{ik}H_{N,kj}$ gives the matrix of the stored orthogonal binary patterns. This can lead to the alternative process of the pattern generation.

Models with planted solutions are an old subject in information theory and statistical physics \cite{chen2022planted,perera2020chook,hen2015probing}. Addressing the planted solution problems appeared in many other domains beyond optimization, e.g. inference-related tasks \cite{zdeborova2016statistical} or image-reconstruction \cite{dong2023phase,maillard2020phase}. For instance, the Wishart planted ensemble was introduced to check the performance of optimization algorithms and related statistical properties \cite{hamze2020wishart}.

\begin{figure}
  \begin{minipage}[c]{0.6\textwidth}
    \includegraphics[width=\linewidth]{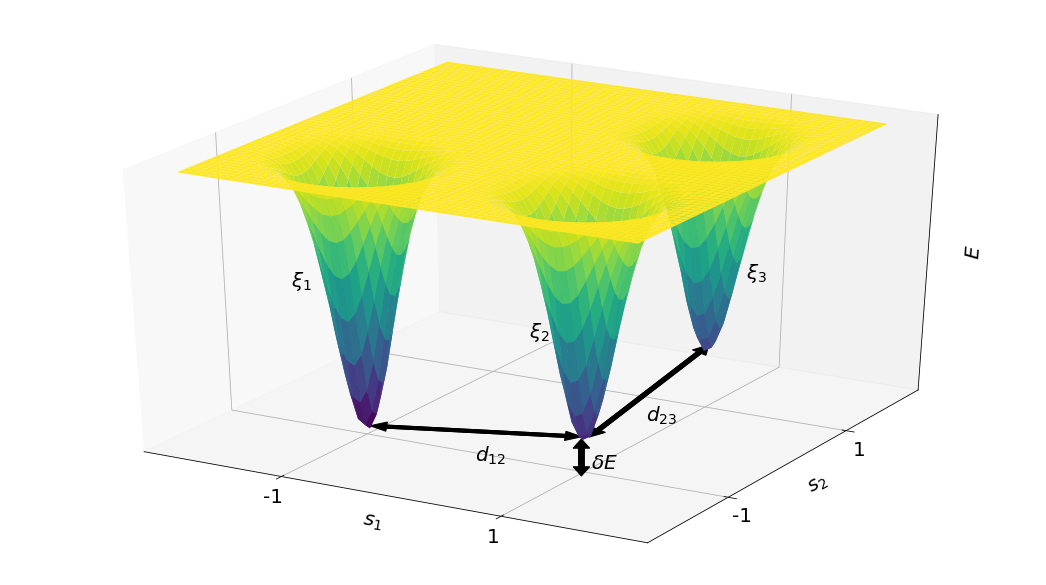}
  \end{minipage}\hfill
  \begin{minipage}[c]{0.4\textwidth}
    \caption{Schematic picture of the modified Hebb's rule. The picture represents the projection of the energy landscape defined by Eq.~(\ref{ising}) with the coupling coefficients given by Eq.~(\ref{modified_hebb}) on the two-dimensional coordinate system for $s_1$ and $s_2$. Both coordinates are considered continuous through the PDE dynamics (e.g. Eq.~(\ref{type1}) or similar equations below) until the projection into the binary states by the $\operatorname{sign}$ function is performed in the end. The energy gap $\delta E$ between the planted patterns $\xi_i$ depends on the parameters $\omega_0$ and $\delta w_i = i \delta w$, while the distance between every pair of minima is chosen to be equal ($N/2$ for even $N$) to preserve their orthogonality. The lower energy minima correspond to the darker colours.
       } \label{schematic}
  \end{minipage}
\end{figure}

We propose the following modification to Hebb's rule:

\begin{linenomath*}
\begin{equation}
J_{i j}=\sum_{\mu \nu}\left(w_0+\delta w_{\mu \nu}\right) \xi_i^\mu\left(Q^{-1}\right)_{\mu \nu} \xi_j^\nu = \sum_{m=1}^K\left(w_0+\delta w_m\right) \xi_i^m \xi_j^m,
\label{modified_hebb}
\end{equation}
\end{linenomath*}

where $\delta w_m = m \delta w$ with $\delta w \ll 1/K$ is the asymmetry coefficient, $K$ is the number of planted patterns, and the patterns have the identity overlap matrix $Q^{-1} = I$ if not mentioned otherwise (except for small-scale instances - see below). Asymmetry perturbations $\delta w$ are chosen to be small in comparison with the $w_0=1$, so they do not affect the patterns basins of attractions but remove their energy degeneracies. For the last pattern $\delta w_K = K \delta w \ll w_0 = 1$. For many reasons, breaking the energy degeneracy with the asymmetry coefficients is essential. Among them is differentiating between the enumerated planted states (each having individual energy), tracking the quality of the final obtained solution, and figuring out the mechanism behind the process of finding the solution. 
Our method still uses the same property of orthogonal patterns, making patterns uniformly distributed across the phase space with equal separation (due to the identity overlap matrix). Another advantage of the orthogonal pattern set is its uniqueness, the possibility of the analytical treatment in the linear regime and explicit information about the ground state energy and energies of the mixed states, which makes it possible to solve the inverse task of reconstructing the initial conditions knowing the final energies. 

The simplest continuous dynamical equation for solving QUBO is known as the Hopfield NN equation \cite{hopfield1982neural,hopfield1985neural} and has the following form (class I):

\begin{linenomath*}
\begin{equation}
\frac{d x_i}{d t}=-\alpha x_i+\sum_j J_{i j} \varphi\left(x_j\right),
\label{type1}
\end{equation}
\end{linenomath*}

with $\alpha$ being the projection strength (also called spherical constraint in the spin-glass community \cite{kosterlitz1976spherical,mezard1987spin,dotsenko1990spin}), $J_{ij}$ are the coupling coefficients of the QUBO problem and $\varphi(x_j)$ is the nonlinear function with the saturation limits and linear behaviour closer to zero values (we use the conventional choice of $\varphi(x_j) = \tanh (x_j)$) and the additional projection into binary values by the end of the simulation. Alternatively, such an equation can be viewed as a combination of the gradient-descent dynamics with the additional projection term specifically for the discrete problems \cite{hopfield1985neural,hertz2018introduction}, the simplest shallow recurrent NN form \cite{hopfield1982neural} or the linearized version of some other dynamical systems \cite{majumdar2014top}. 

Thus, the description of the dynamical equation given by Eq.~(\ref{type1}) on the energy surface given by Eq.~(\ref{modified_hebb}) is straightforward: one starts with the random (with no prior information) coordinates, which corresponds to either the basin of the planted pattern or spurious state, and then follows the gradient descent until reaching the steady state and consequent projection into corners of the hypercube. The steady state is achieved by compromising between the gradient-descent and slow projection terms (or any other terms, depending on the details of the dynamical equation). Our method distinguishes itself by individually treating the planted minima by separating their (and corresponding mixed states) energies and an additional detailed evaluation of the different dynamical equations' performance on the produced instances. 
The schematic picture of this idea and the corresponding energy landscape is shown in Fig.~{\ref{schematic}}.

We present the calculations on different scales and with the different classes of dynamical equations according to our classification, ranging from the simple Hopfield NN to the complicated dynamics with higher derivatives and annealing schedules, part of which are presented in the Supplementary Material section. However, much emphasis will be put on the simplest Eq.~(\ref{type1}). Additional information about the classification of dynamical equations and their transformations with the corresponding references are presented in \cite{syed2022physics,stroev2023analog}. The Class I dynamical equation in our notation is every partial differential equation that possesses a simple form and can be mathematically reduced to the Eq.~(\ref{type1}). However, in the presence of variations, such as time dependency of $\alpha(t)$ or the presence of an additional term with a higher derivative, the equation no longer falls into the Class I category. In such cases, it becomes necessary to reclassify it into other classes.

\section{Small scale classification}
\label{Small scale classification}

\begin{figure}
  \begin{minipage}[c]{0.5\textwidth}
    \includegraphics[width=\linewidth]{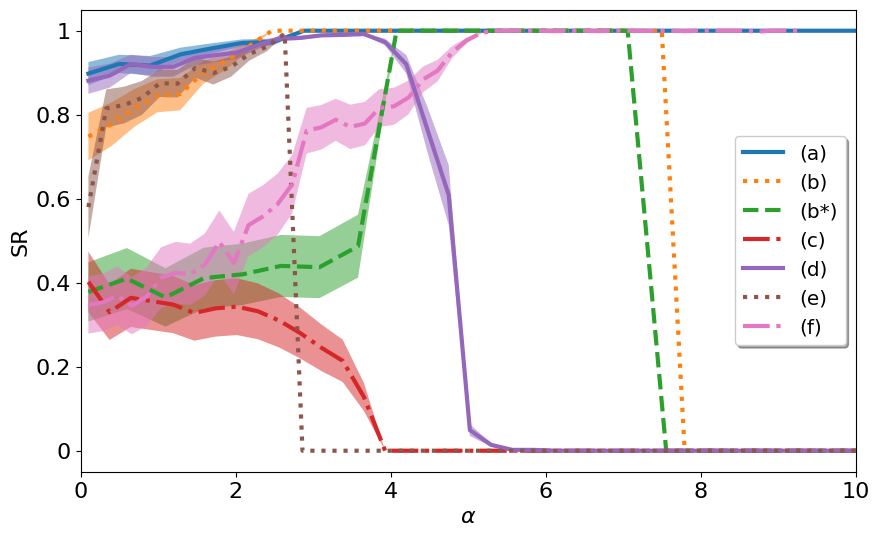}
  \end{minipage}\hfill
  \begin{minipage}[c]{0.4\textwidth}
    \caption{Success rate (SR) in achieving the ground state for the low-scale matrix models for $\beta=1$ over various $\alpha$ values. For all instances we have SR$=1$ except for the matrix (c) when evolved with Eq.~(\ref{type1}). The range of $\alpha$ with SR$=1$ depends on the patterns' overlap. The shaded zones represent possible errors in evaluating the SR due to finite sampling of initial conditions. (a), (b), etc. correspond to the coupling matrices discussed in the main text.
       } \label{abcdef_profiles}
  \end{minipage}
\end{figure}

Before describing the construction of the problem instances, it is crucial to make a few comments about the symmetries in our system, which take only quadratic interactions (so-called zero-field Ising Hamiltonian). According to Hebb's rule, planting a pattern $\xi_i$ will produce its mirror "twin" $-\xi_i$ in the phase space, which reflects the mirror symmetry. Thus, changing a sign of all components of some of the planted pattern will not affect the energy landscape. Moreover, the system has a gauge symmetry, which allows one to reverse particular spins in planted patterns without affecting the complexity of the energy landscape. Nevertheless, such transformations will manifest themselves in the $J_{ij}$ coefficients.

Figure {\ref{abcdef_profiles}} compiles nontrivial Success Rate (SR), which is the ratio of the runs that reached the ground state to all runs by a particular algorithm for a given instance, profiles across all $N=8$ and $K=3$ matrix instances, and captures diverse dynamic behavior of Eq.(\ref{type1}) for finding minimal energy solutions on these matrices. Below, we emphasize the essential role of matrix choice, explore inter-matrix differences, delve into parametrization details, and outline our complexity framework.

We introduce a few small-scale matrices with the corresponding classification on which the presented dynamical systems were tested and showed various behaviours. For such size, the essential parameters are the number of planted patterns $K$ ($K=3$ in the majority of the cases, so that it is easy to observe the planted patterns and their spurious states without additional effects), Hamming distances between the discrete patterns (or their overlaps), and the value of the asymmetry coefficient $\delta w$, which should be small enough not to affect the patterns basins of attraction. In case of large pattern overlaps, the patterns tend to merge into one global minimum, and the same effect can be achieved by increasing the asymmetry coefficient. Thus, it is reasonable to classify instances by the triplet of the mutual distances between the planted patterns $(d_{12},d_{23},d_{13})$ (where we used conventional Hamming distance as the measure), which produce several possible combinations, the number of which is constrained by the triangle inequality and possible distances $d_{ij}<5$, because planting the pattern comes together with its reverse copy in the absence of the external field.
For example, two patterns $(1,1,1,1,1,1,1,1)$ and $(1,1,1,1,-1,-1,-1,-1)$ have the Hamming distance of $4$.
The list of nontrivial configurations is following: a) $(1,3,4)$ with $\delta w_a =\delta w = 0.1$, b) $(4,3,3)$ with $\delta w_b = 0.1$, b$^{*}$) $(2,4,2)$ with $\delta w_{b^{*}} = 0.3$, c) $(3,3,4)$ with $\delta w_c = 0.1$, d) $(4,4,2)$ with $\delta w_d = 0.1$, e) $(4,1,3)$ with $\delta w_{e} = -0.13$ (with the $3^{d}$ pattern having the biggest weight - $w_0 + 3 \delta w$ accroding to the Eq.~(\ref{modified_hebb})) and specific exceptional example with four patterns f) $(4,4,4,4)$ with four equidistant vectors and $\delta w_f = 0.1$ (to test the complexity beyond the given classification). 
Nontrivial configurations are configurations that produce unique SR profiles over $\alpha$ (considering such features as the SR values for large and small $\alpha$, existence of the $\text{SR}=1$ plateau or its absence and the monotonic behavior for intermediate SR values). Other configurations either possess trivial properties (for example the set of distances $(1,2,1)$ produces one merged global minima)) or produce the SR profiles that all fall into some of these functions.
From the geometrical perspectives, the set of patterns distances uniquely defines the basins of attraction for planted patterns and their mixed combinations and hence the energy landscape.
Several reasons make instances of this kind ideal for testing algorithms and hardware. Firstly, they exhibit few basins of attraction, making them easy to characterize. Additionally, success rate diagrams can be obtained using alternative methods, enabling comparisons to identify performance differences across methods and parameter variations. Any deviations observed in these diagrams can unveil signatures of differences in various optimization mechanisms. Moreover, these artificial examples play a crucial role as they constitute the initial testing stage for numerous experimental works before scaling up the experimental setup in subsequent phase.

In case of the limited capabilities or constraints imposed by the analogue nature of the hardware, the produced matrices can be coarse-grained into another set by the following transformation $\widetilde{J_{ij}} = \left\lfloor J_{ij} / \Delta J \right\rfloor$ (depending on the reasonable choice of the discretization parameter $\Delta J$, $\left\lfloor x \right\rfloor$ denotes the floor function that outputs the greatest integer less than or equal to $x$) and preserve the same geometrical properties of the energy landscape and thus complexity (which was verified with the corresponding numerical simulations). One can tune the set by changing the $\Delta J$ parameter, which depends on the analogue hardware properties and its working precision \cite{mourgias2023analog}.



The numerical evolution of Eq.~(\ref{type1}) on the given instances gives different results. It was performed using the simple Euler scheme with $N_{\text{iter}}=1000$ iterations of $dt=0.01$ timesteps, with $100$ samples per parameter $(\alpha_j,\beta_j)$ point.For small-scale instances, we previously defined the complexity through SR (the ratio of the runs that reached the ground state to all runs by a particular algorithm for a given instance). The absence of the region with SR$=1$ is treated as a hard instance, which is the case of the (c) matrix. Easier matrices have all a larger region with SR=$1$ or larger SR over all parameters. In the case of (c) instance, we want to effectively "hide" the ground state by placing other patterns closer to each other to merge into another minimum, which still does not exceed the ground state energy. By prioritizing planted patterns with high eigenvalues, one can exploit the mathematical framework to find the global minimum. Removing most eigenvectors and eigenvalues, it is possible to use power iterations to approach the ground state. To prevent straightforward exploitation, our goal is effective "hiding" of the ground state by introducing correlations among local minima, complicating the search and rendering the power method ineffective. As a result, the system's dynamic behavior aligns with the coupling matrix's highest eigenvalue eigenvector, arising from the interaction of the second and third patterns. This vector differs from the ground state pattern, making it nearly impossible to reach the lowest energy from any point in the phase space except for initial conditions very close to it.
Using Eq.~(\ref{modified_hebb}) for defining the energy landscape, we can guarantee that the ground state corresponds to the pattern with the highest asymmetry coefficient, which is the principle eigenvalue of the $J$. Raising the parameter $\alpha$ effectively raises the threshold for eigenvalues. As a result, during the dynamical evolution, the eigenvectors corresponding to eigenvalues below this threshold will undergo an exponential reduction in size due to consecutive matrix-vector multiplications. Increasing $\alpha$ further turns the dynamical equations into a trivial fast projection of the initial state, which does not correlate with the ground state. The detailed SR diagrams for all instances (except (e), which is close to the (b) matrix) and statistics for other dynamical equations can be found in the Supplementary Material.

\begin{figure} 
\centering
\begin{tabular}{cccc}
\includegraphics[width=0.3\textwidth]{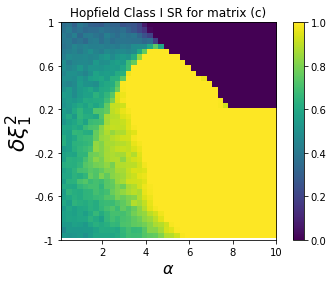} &
\includegraphics[width=0.3\textwidth]{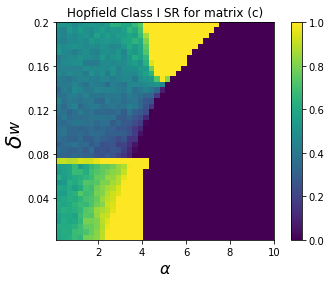} &
\includegraphics[width=0.3\textwidth]{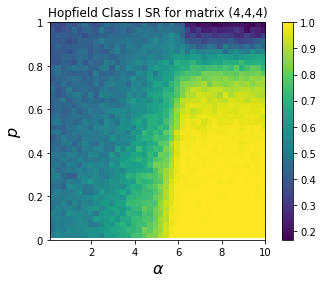} 
\end{tabular}
\caption{Various complexity transition diagrams for the SR according to Eq.~(\ref{modified_hebb_2}) parametrization. Left: Countor plot of SR dependence on the spin value (coinciding with the value of other patterns) for the pattern with the lowest weight. Middle: SR dependence on the asymmetry coefficient in the modified Hebb's rule versus the projection strength. Right: SR dependence on the pattern parametrization through the parameter $p$; see the main text for details of the parametrization.
}
\label{complexity_transitions}
\end{figure}

Another noteworthy aspect involves the emergence of complexity transitions within a singular model, that are shown in Fig.~{\ref{complexity_transitions}}. Our approach entails the exploration of diverse parameterizations for the model, as defined by the equation Eq.~(\ref{modified_hebb}). This parameterization has the following form:

\begin{linenomath*}
\begin{equation}
J_{i j}=\sum_{m=1}^K\left(w_0+\delta w_m\right) (\xi_i^m + \delta \xi_i^m) (\xi_j^m + \delta \xi_j^m),
\label{modified_hebb_2}
\end{equation}
\end{linenomath*}

We start with the type (c) matrix since it has a complex profile without reaching $100 \%$ SR. 

The left part of Fig.~{\ref{complexity_transitions}} shows the transition between the states, where we continuously changed the second coordinate of the first pattern in (c) matrix (or any other coordinate which shares a sign across planted patterns) from its original value $\xi_{1}^{2}=1$ to the opposite $-1$ (which corresponds to the $\delta \xi_{1}^{2}$ going from $0$ to $-2$), while keeping other spins constant. 
Such a procedure effectively changes the distances between patterns, which affects $J_{ij}$ coefficients and hence the complexity of the current instance. As seen in the left part of Fig.~{\ref{complexity_transitions}}, the first transition happens with the appearance of the yellow region, which makes the generated instance easy ($\delta \xi_{1}^{2} \sim 0.7 $), while the second is marked by the disappearance of the region with the $0 \%$ SR ($\delta \xi_{1}^{2} \sim 0.2 $). In our context, we call the instance "easy" if its SR diagram possesses region with SR$=1$. The middle part of Fig.~{\ref{complexity_transitions}} shows the same (c) matrix SR diagram but with different asymmetry coefficients. Increasing $\delta w$ from its original value $\delta w=0.1$ leads to the dominance of the pattern with the largest weight. Decreasing it also leads to complexity reduction with sharp boundaries.

The right part of Fig.~{\ref{complexity_transitions}} shows the diagram for the specific modification of the equidistant planted patterns with $\xi_{1}^{4}=\xi_{2}^{4}=1+0.2p$ and $\xi_{3}^{4}=-1+0.1p$ (index $4$ stands for the spin that has a common sign for first two patterns), which becomes hard through the described transition. Even with such a small shift, it is possible to obtain a complexity transition, imposed by current parametrization. However, the boundaries are rather smooth, in contrast to the previously described cases. 
In summary, we introduced various instances of coupling matrices $J_{ij}$ for the QUBO problem of size $N=8$ and systemized them based on the mutual distances. We then demonstrated various ways to modify these instances affecting their complexity. These methods can be used for larger matrices.


\section{Numerical results for large N}
\label{Numerical results for large N}

The transition to the larger $N$ allows one to plant more patterns, which leads to more sophisticated scenarios and effects. However, one has to modify the criteria for the hardness of a matrix because of the increasing difficulty of finding the true ground state across all the possible solutions dominated by spurious patterns. Thus, we shift our attention to other characteristics, like the functional form of the final solutions distribution, its dependency on the number of planted patterns, and the distribution of found energies relative to the planted ones.


\begin{figure}
\begin{minipage}[c]{0.95\linewidth}
\includegraphics[width=\linewidth]{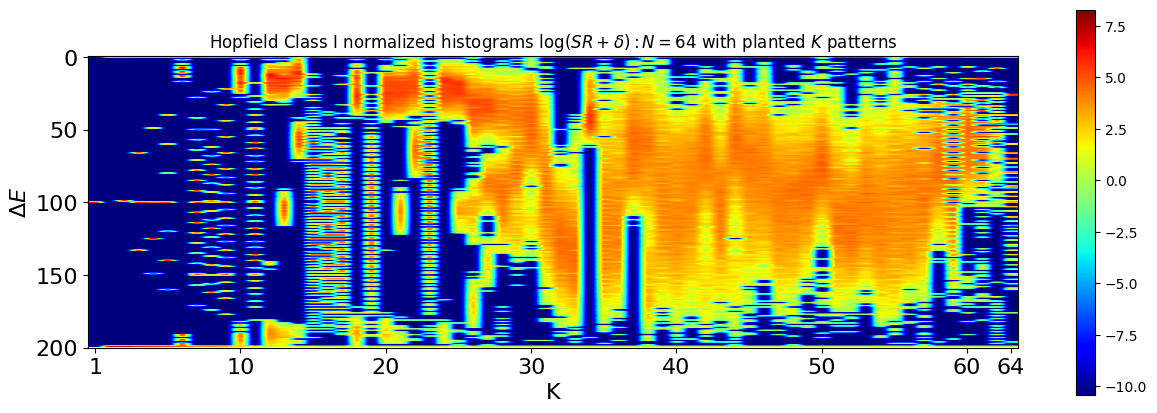}
\caption{ The $\log$ distribution of the found energies (smoothed histograms), centred between the found maximal $\widetilde{E_{max}}$ and minimal $\widetilde{E_{min}}$ energy values, obtained after a series of simulations of Eq.~(\ref{type1}) on the medium-scale problems. A log scale is chosen to highlight the distribution shape. $\Delta E$ denotes the introduced discretization of the $[\widetilde{E_{min}},\widetilde{E_{max}}]$ range. There are a few notable regimes, depending on the number of planted patterns $K$ described in the text. $\delta = 0.00003$ is a small parameter chosen to shift the onset of the logarithm from the infinitely small values.
}
\label{histograms_shape}
\end{minipage}
\begin{minipage}[c]{0.95\linewidth}
\includegraphics[width=\linewidth]{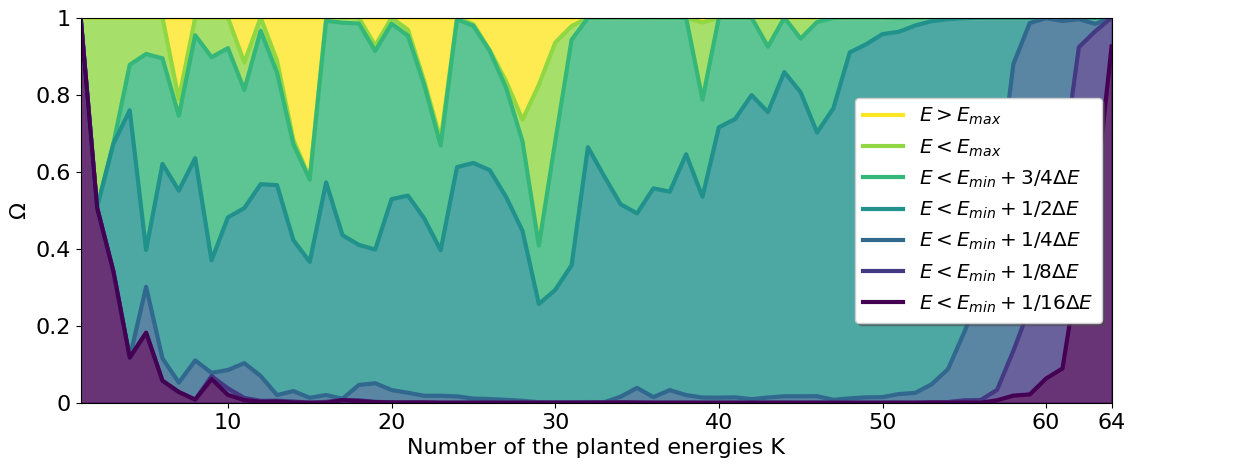}
\caption{ The relative frequency distribution of the energies of discovered solutions in relation to the planted energies. $\Omega$ presents the entire probability space, which is partitioned based on the likelihood of events associated with finding a solution within specified energy ranges (denoted by the corresponding color). Different fractions ($1/16,1/8..$ etc.) are chosen to separate the $[E_{min},E_{max}]$ range. Here, $[E_{min},E_{max}]$ are defined as the planted minimal and maximal energies related to the individual patterns, not found ones. The corresponding description of various scenarios depending on the $K$ and the dominated energies can be found in the main text.
}
\label{histograms_measure}
\end{minipage}
\hfill
\end{figure}

Dimension $N = 64$ produces quite a reasonable space for planting many patterns, because it has enough capacity for the orthogonal patterns to produce mixed states that challenge the sampling of points with energies near the ground state. We perform extensive numerical simulations of Eq.~(\ref{type1}) over the instances defined by the coupling matrices of Eq.~(\ref{modified_hebb}) for each $K=1..N$ with $\alpha = \lambda_{max}/2$ (where $\lambda_{max}$ is the principal eigenvalue of the $J_{ij}$ coupling matrix) until the steady state is found. For initial conditions, we use vectors with independent random coordinates uniformly distributed in the interval $x_i \in [ -a, a]$, with small enough $a = 0.5$. The numerical integration was performed using the simple Euler scheme with $N_{\text{iter}}=1000$ iterations of $dt=0.01$ timesteps. Although it is costly to characterise the entire phase space, we perform an appropriate sampling, enough to characterize the energy landscape (in terms of the energy (and hence the relative position with respect to the planted energies) and the coefficients in the planted patterns decomposition).
After reaching the steady state, we calculate the energy according to Eq.~(\ref{ising}), normalise it using the maximal $\widetilde{E_{\max}}$ and minimal $\widetilde{E_{\min}}$ found energy values and plot in the Fig.~{\ref{histograms_shape}}.

To highlight the shape of the empirical energy distribution, Fig.~{\ref{histograms_shape}} shows the smoothed histograms centred between the maximal $\widetilde{E_{\max}}$ and minimal $\widetilde{E_{\min}}$ empirically found energy values, demonstrating a variety of scenarios. There are few notable regimes, depending on the number of the planted patterns $K$. For small $K\lesssim 10$, there are separated planted patterns with their mirror copies and spurious states. For moderate $K$: $N/2 \gtrsim K\gtrsim 10$, the number of found spurious states becomes bigger, resulting in the cluster's appearance and multimodal distributions, sometimes mixed with the sparse histograms of the same spurious states. For large $K$ the dominance of the spurious states becomes evident, resulting in the normal distributions reaching the saturated regime $K\gtrsim 58$, where the distributions become trivial again. Such trivialization can be explained. Taking equal weights as in the original Hebb's rule and placing the $K=N$ orthogonalized patterns will produce the matrix with zero off-diagonal elements \cite{hertz2018introduction}. Thus, taking $K=N$ in Eq.~(\ref{modified_hebb}) effectively renormalizes the weights into $\delta w_m$, because terms with the $\omega_{0}$ cancel each other, leaving weights with linear dependency on $m$, which has an exponential impact on the dynamics of convergence to a particular pattern. The complexity of finding the lowest or sub-optimal energy state depends on the complexity parameters: $K$, which regulates the number of local minima, patterns distances between each other (with the orthogonal conditions being the most complex) and asymmetry coefficients $\delta w_m$ (which we do not focus on).

Figure~{\ref{histograms_measure}} illustrates the relative frequency distribution of the energies of discovered solutions in relation to the planted energies. The symbol $\Omega$ represents the entire probability space, which is partitioned based on the likelihood of events associated with finding a solution within specified energy ranges. The probability of finding patterns with low enough energies relative to the planted energies is comparatively low in the primary region $10<K<55$, where most obtained energies are localized below the average value of the planted energies (red colour). $0.5 \Delta E$ is the actual average of the planted energies and it depends on the parameter K. Another significant measure consists of patterns above the mean but below three-quarters of the planted energies range. There are found solutions that lie outside of the given range for some $K$. For other values of $K<10$ or $K>55$, the dynamics are trivial, and the solutions are closer to the ground state. A more detailed picture can be found in the Supplementary Material.

\begin{figure}
\begin{minipage}[c]{0.95\linewidth}
\includegraphics[width=\linewidth]{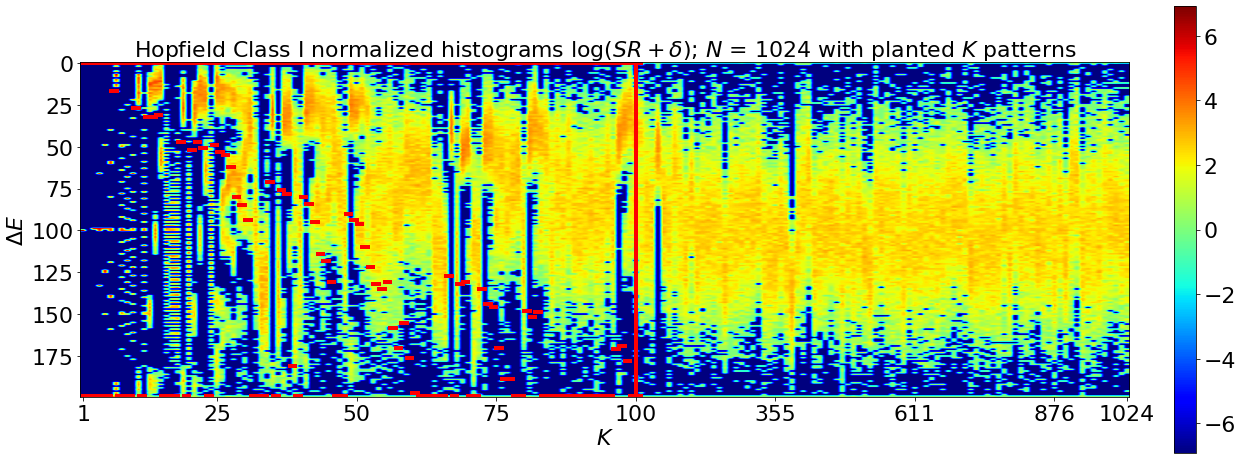}
\caption{
The $\log$ distribution of the found energies (smoothed histograms), centred between the found maximal $\widetilde{E_{max}}$ and minimal $\widetilde{E_{min}}$ energy values, obtained after a series of simulations of Eq.~(\ref{type1}) on the large-scale problems. A log scale is chosen to highlight the distribution form better. $\Delta E$ denotes the discretization of the $[\widetilde{E_{min}},\widetilde{E_{max}}]$ range. There are a few notable regimes, depending on the number of planted patterns $K$ described in the text and the simulation details. Red marks denote the range of the planted energies. The right half of the diagram is compressed $\approx 9$ times because of the dominated Gaussian-like distributions. $\delta = 0.00003$ is a small parameter chosen to shift the onset of the logarithm from the infinitely small values.}
\label{histograms_shape_1024}
\end{minipage}
\begin{minipage}[c]{0.95\linewidth}
\includegraphics[width=\linewidth]{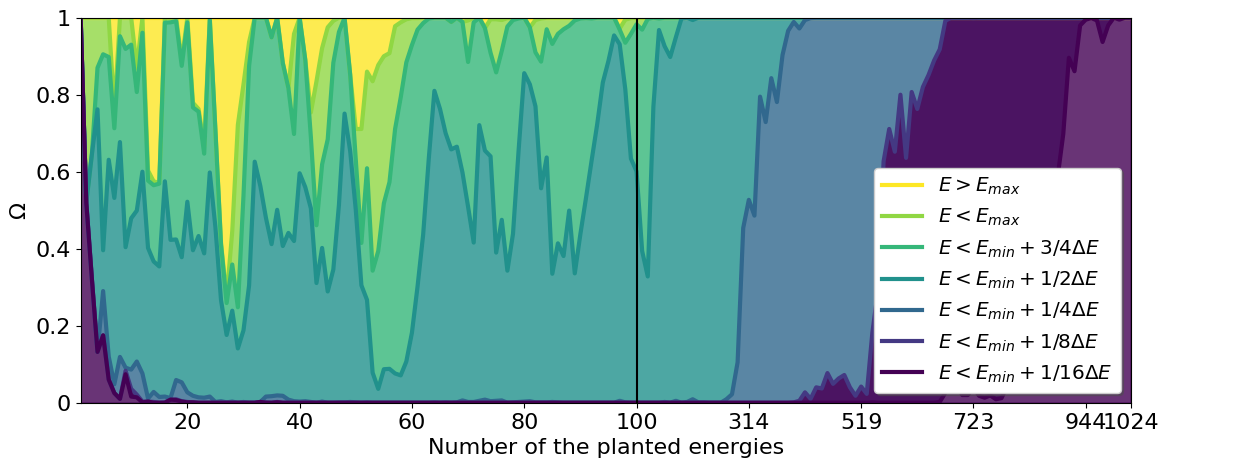}
\caption{The relative frequency distribution of the energies of discovered solutions in relation to the planted energies. $\Omega$ presents the entire probability space, which is partitioned based on the likelihood of events associated with finding a solution within specified energy ranges (denoted by the corresponding color). Different fractions ($1/16,1/8.,$ etc.) are chosen to separate the $[E_{min},E_{max}]$ range. The corresponding description of various scenarios depending on the $K$ and the dominated energies can be found in the main text. The region of $K>100$ can be used as a statistical test due to the localisation of the found solutions.}
\label{histograms_measure_1024}
\end{minipage}
\hfill
\end{figure}

Finally, we perform similar numerical simulations for $N = 1024$ over the instances defined by Eq.~(\ref{modified_hebb}) for each $K=1..N$ with the projection strength equals half of the maximal eigenvalue of the matrix $\lambda_{max}/2$ and same initial conditions as in $N=64$ case. The numerical integration was performed using the simple Euler scheme with $N_{\text{iter}}=1000$ iterations of $dt=0.0005$ timesteps. We perform an appropriate sampling, enough to evaluate various characteristics and their measure, although the amount of runs is less than in the previous case. 

To present the distribution form, Fig.~{\ref{histograms_shape_1024}} shows the histograms centred between the found maximal and minimal energy values. We concentrated on the same indicators as in the previous $N=64$ case. Many patterns appearing, in this case, are similar. For the $K \geq 100$, the dominance of the spurious states results in close to normal distributions (until the saturated regime). However, such a regime occupies a bigger range of $K$. Therefore, we expect the ratio of the $K/N$ for the starting value of the corresponding regime to be lower with the increase in $N$. For small $K\lesssim 25$, there are well separated planted patterns with their mirror copies and spurious states. After $K\gtrsim 25$, there are mostly cluster and multimodal distributions of minima, which slowly transform close to normal distribution.

Figure ~{\ref{histograms_measure_1024}} illustrates the relative frequency distribution of the energies of discovered solutions in relation to the planted energies. In the region $20<K<100$, most of the obtained energies are localized below the energies mean (red colour), except for some particular regions, and the distributions are more diverse than in the $N=64$ case. For the higher value of $K>100$, one can use good approximations for the Gaussian distributions with $\mathbb{E}[E] \approx E_{min} + \Delta E \cdot 2^{-1-K/200} $. Choosing $200<K<800$ for constructing the matrix instance will possess a problem of reasonable complexity for linear gradient-descent-like or sampling algorithms. The most difficult region is situated approximately at the $K \approx 500$ with a slight shift to the right. A more detailed picture can be found in the Supplementary Material Section together with the numerical results with other algorithms and the exploration of the challenging regime.

The asymmetry parameters $\delta w = 0.1$ for $N=8$ and $\delta w = 0.001$ for $N=64$ and $N=1024$ were picked intentionally small not to affect the basins of attraction. However, this choice leads to a slight numerical difference between found solutions, and one must be careful to differentiate them properly. The choice of the projection strength equals half of the maximal eigenvalue $\alpha = \lambda_{max}/2$ allows one to have a reasonable convergence as well as to obtain a sample of the close to the optimal solutions with the empirical evidence (see the left picture in Fig.~{\ref{n3c_diagrams}}). However, the projection strength significantly affects the form of the distribution; see the additional plots in the Supplementary Material section.

\section{Discussion and conclusions}
\label{Discussion and conclusions}

We introduced universal, easy-to-reproduce generative models for the QUBO instances to differentiate the performance of the hardware/solvers. One can use different aspects of such benchmarks. By tuning the complexity, it is possible to determine the performance of the algorithm/hardware on the given instance. Different methods can be compared and contrasted by their probability of reaching the ground state or particular energy interval of sub-optimal solutions. One can also measure the output energy distributions and compare them with the obtained by current numerics. The precise correspondence or mismatch between found energies is an indicator of the difference in methods descriptions or physical principles of unconventional computing. In summary, one can use the whole model, its particular instances or some reproduced patterns with various parameters to compare the hardware performance. We investigated small, medium and large-scale problems and their properties. The described approach is easy to implement and can be successfully applied to benchmark classical and quantum devices. Moreover, our method establishes a prospective platform to characterize the performance of the hardware devices' physical processes and algorithms for novel computing machines.

Compared to the conventional random matrix ensembles, which are also widely used to perform the optimization, our model has many more local minima and offers higher complexity in finding the optimal or sub-optimal solution (assuming the the method does not incorporate any prior information about the model). Another advantage is the classification of possible states and trajectories, which is statistically hard to describe for the fully random instances.

The presented approach lies at the intersection of many fields, such as tensor factorization (since Eq.~(\ref{hebb}) is essentially rank decomposition of a matrix) \cite{cichocki2009nonnegative} or statistical physics and dynamics of complex systems \cite{kadanoff2000statistical}, not to mention that Eq.~(\ref{type1}) (and its modifications) is the Hopfield equations \cite{hopfield1982neural,hopfield1985neural}, the simplest form of the NN. 


\section{Supplementary Material}
\label{Supplementary Material}


\renewcommand\thefigure{S\arabic{figure}}
\setcounter{figure}{0}  

This Section contains supplementary information and additional numerical results. We present the calculations on different scales and with the different classes of dynamical equations according to our classification, which is presented below, ranging from the simple Hopfield NN to the complicated dynamics with higher derivatives and annealing schedules. Class II equations differ from the previous class by the presence of time dependence in its coefficients:

\begin{linenomath*}
\begin{equation}
\frac{d x_i}{d t}=-\alpha(t) x_i+\beta(t) \sum_j J_{i j} \varphi\left(x_j\right).
\label{type2}
\end{equation}
\end{linenomath*}

The class III equations have the following form:

\begin{linenomath*}
\begin{equation}
\frac{d^2 x_i}{d t^2}=\gamma(t) \frac{d x_i}{d t}-\alpha(t) x_i+\beta(t) \sum_j J_{i j} \varphi\left(x_j\right),
\label{type3}
\end{equation}
\end{linenomath*}

and are governed by second-order partial differential equations with time-dependent coefficients, which is the special case of the coupled microelectromechanical systems (MEMs) equations \cite{hoppensteadt2001synchronization}. Similar equations describe bifurcation machines \cite{goto2019combinatorial,goto2021high,kanao2022simulated}:

\begin{linenomath*}
\begin{equation}
\frac{d^2 x_i}{d t^2}=-\Delta(\Delta-p) x+\Delta \xi_0 \sum_j J_{i j} \varphi\left(x_j\right),
\label{TBM}
\end{equation}
\end{linenomath*}

where $\Delta$ is the positive detuning frequency between the oscillator and pumping frequency, $p$ is the external pumping parametrization and $\xi_{0}$ is a positive constant with the dimension of frequency from the original model \cite{goto2019combinatorial}.
The clear analogy between different parametrization of Eq.~(\ref{TBM}) and Eq.~(\ref{type3}) is following $\alpha = \Delta (\Delta - p)$ and $\beta = \Delta \xi_0 $ (with $\gamma = 0$). However, all the equations describing various dynamical rules have many connections, with possible transformations between each other \cite{syed2022physics,stroev2023analog}.

\begin{figure} [H]
\centering
\begin{tabular}{cccc}
\includegraphics[width=0.3\textwidth]{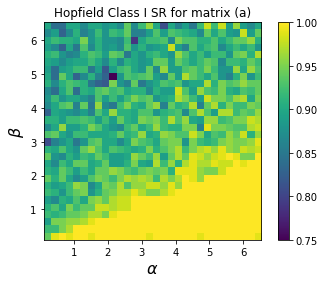} &
\includegraphics[width=0.3\textwidth]{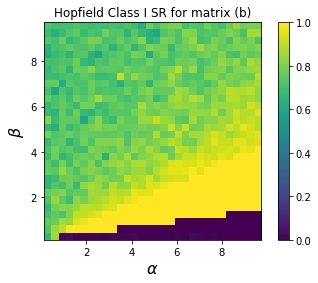} &
\includegraphics[width=0.3\textwidth]{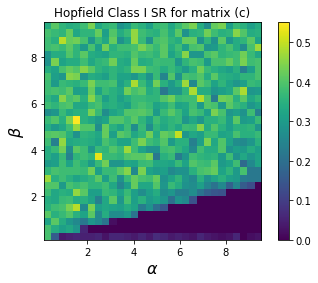} 
\end{tabular}
\caption{Success rate (SR) of finding the global minimum of the low-scale QUBO models. SR depends only on the ratio of $\alpha/ \beta$, where $\beta$ is the prefactor before the coupling coefficients $J_{ij}$ in Eq.~(\ref{type1}). Matrix (a) with distances $(1,3,4)$ has the region with SR $=1$ for higher projection strengths. For matrix (b) $(4,3,3)$ higher $\alpha$ results in the projection that always leads to a suboptimal solution. Matrix (c) $(3,3,4)$ with $dw = 0.1$ leads to the most challenging instance manifested by the lack of region with SR$=1$. Overlap of the planted patterns significantly influences SR, which is why the (c) instance allows one to hide the ground state, which does not coincide with the maximal eigenvector. The details of the numerical integration of the dynamical equations do not affect SR. The nature of speckles depends on the finite amount of samples for the averaging procedure and should disappear if averaging over many simulations is made.
}
\label{abc_diagrams}
\end{figure}

\begin{figure} [H]
\centering
\begin{tabular}{cccc}
\includegraphics[width=0.3\textwidth]{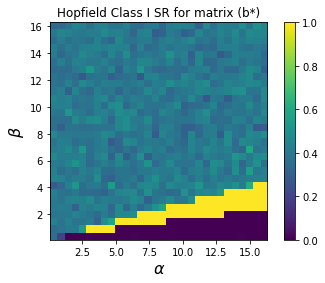} &
\includegraphics[width=0.3\textwidth]{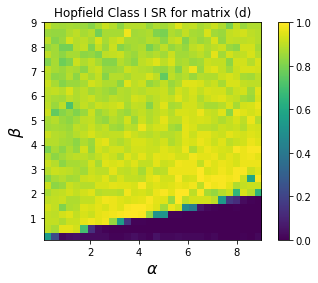} &
\includegraphics[width=0.3\textwidth]{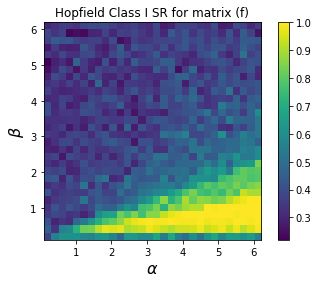} 
\end{tabular}
\caption{Success rate (SR) of finding the global minimum of the low-scale QUBO models. Matrix (b$^{*}$) with distances $(2,4,2)$ and $dw_{b^{*}} = 0.3$ has the region with SR$=1$ with very sharp bounds. For matrix (d) $(4,4,2)$ and $dw = 0.1$, there is no region with SR$=1$, despite the probability of reaching the ground state being relatively high. Matrix (f) $(4,4,4,4)$ with $dw = 0.1$ leads to a different SR diagram compared to other cases, as it lacks a region where SR=$0$. The details of the numerical integration of the dynamical equations do not affect SR.
}
\label{fbd_diagrams}
\end{figure}

Although for the class I algorithm SR depends only on $\alpha/ \beta$ (or just $\alpha$, assuming $\beta=1$), we plot an additional axis $\beta$ to ensure no numerical mistake. Thus SR is constant across any linear ratio $\alpha / \beta$, see Fig.~{\ref{abc_diagrams}} and Fig.~{\ref{fbd_diagrams}}. 

\begin{figure} [H]
\centering
\begin{tabular}{cccc}
\includegraphics[width=0.3\textwidth]{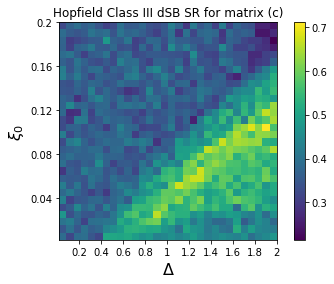} &
\includegraphics[width=0.3\textwidth]{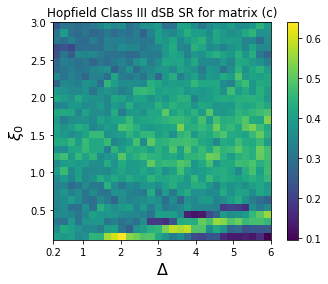} &
\includegraphics[width=0.3\textwidth]{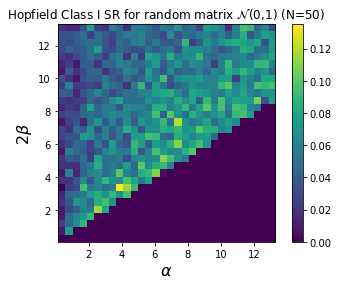} 
\end{tabular}
\caption{Left and middle: success rate (SR) of finding the global minimum of the QUBO with the (c) coupling matrix using Eq.~(\ref{TBM}) with $\varphi (x_j) = \text{sign}(x_j)$ and derivative $d x_j/ d t$ existing for the argument $x_j$ only in the $[-1,1]$ region. Numerical integration was performed with $N_{t} =1000$ timesteps of $dt=0.1$ using the Euler scheme and parameter annealing in the form $p = \text{min}(2t/N_{t},2)$. The right plot shows the performance of Eq.~(\ref{type1}) on the average instance of the symmetric random matrix of size $N=50$.}
\label{n3c_diagrams}
\end{figure}

Additional numerical results include left and middle plots in Fig.~{\ref{n3c_diagrams}} that shows the gain in the SR by a transition to more complicated dynamics of Eq.~(\ref{TBM}). The right plot in Fig.~{\ref{n3c_diagrams}} shows the dynamics of Eq.~(\ref{type1}) algorithm on the average instance of a symmetric random matrix of size $N=50$, which typically does not have a region with the SR$=1$. This diagram can guide the choice of optimal projection strength $\alpha$.

\begin{figure} [H]
\centering
\begin{tabular}{cccc}
\includegraphics[width=0.3\textwidth]{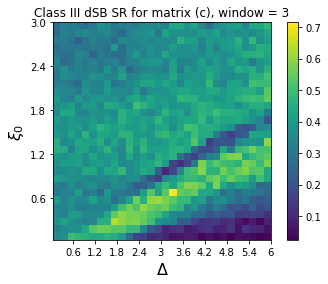} &
\includegraphics[width=0.3\textwidth]{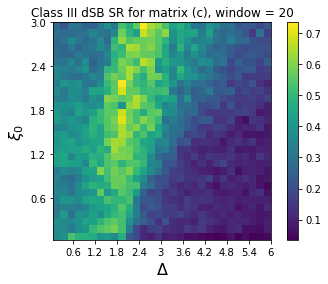} &
\includegraphics[width=0.3\textwidth]{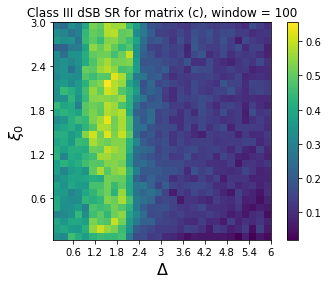} 
\end{tabular}
\caption{Success rate (SR) of finding the global minimum of the QUBO with the (c) coupling matrix using Eq.~(\ref{TBM}) with $\varphi (x_j) = \text{sign}(x_j)$ and derivative $d x_j/ d t$ existing for the argument $x_j$ only in the $[-1,1]$ region. Numerical integration was performed with $N_{t} =1000$ timesteps of $dt=0.1$ using the Euler scheme and parameter annealing in the form $p = \text{min}(2t/N_{t},2)$. The window serves as the threshold parameter (default value $=1$), scaling the region where the derivative $d x_j/ dt$ is considered to exist.}
\label{window_diagrams}
\end{figure}

Fig.~{\ref{window_diagrams}} shows the change in the SR diagrams for Eq.~(\ref{type3}) algorithm on the (c) matrix depending on the choice of the threshold argument value.

\begin{figure}[H]
\begin{minipage}[c]{0.95\linewidth}
\includegraphics[width=\linewidth]{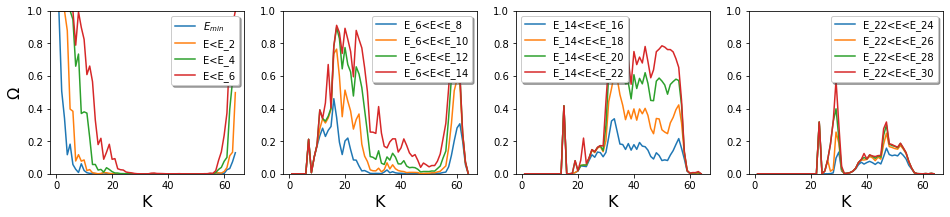}
\caption{ The relative frequency distribution of the energies of discovered solutions in relation to the planted energies showing individual energies up to the $30$ patterns for the $N=64$. 
For initial conditions, we use vectors with the coordinates being independent random variables uniformly distributed in the interval $x_i \in [ -a, a]$, with small enough $a = 0.5$. $\alpha = \lambda_{max}/2$ 
}
\label{histograms_measure_cuts}
\end{minipage}
\hfill
\end{figure}

\begin{figure}[H]
\begin{minipage}[c]{0.95\linewidth}
\includegraphics[width=\linewidth]{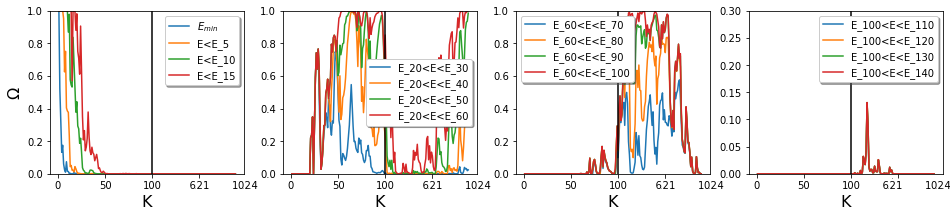}
\caption{The relative frequency distribution of the energies of discovered solutions in relation to the planted energies with the higher discretization up to the $140$ patterns for the $N=1024$. }
\label{histograms_measure_cuts_1024}
\end{minipage}
\hfill
\end{figure}

Figs.~{\ref{histograms_measure_cuts}} and {\ref{histograms_measure_cuts_1024}} focus on a more detailed picture of the results, presented in the main text, providing additional information about the measure of the particular solutions found.

\begin{figure}[H]
\begin{minipage}[c]{0.95\linewidth}
\includegraphics[width=\linewidth]{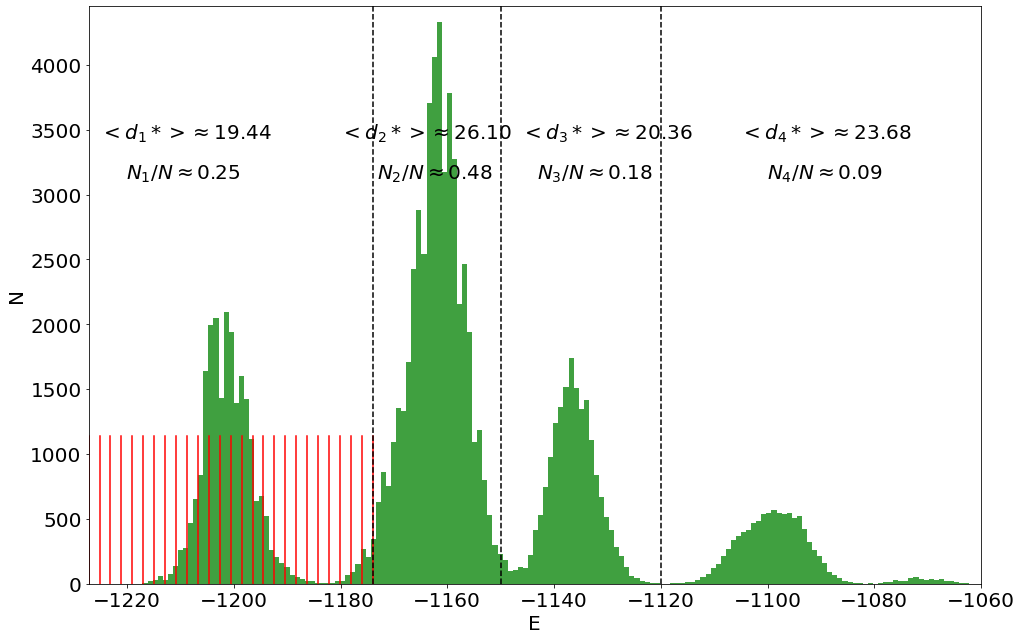}
\caption{Energy histograms of solutions of Eq.(\ref{type1}) for $K=26$ out of the $N=64$ for $\alpha = \lambda_{max}$, demonstrating the multimodal distribution, which is the result of the interplay between GD-dynamics and discrete projection. We informally separate the distribution into the four clusters, characterized by the overall measure $N_i/N$ and the average Hamming distance between solutions inside a cluster $<d_i *>$. Red lines denote the planted energies.
}
\label{multimodal}
\end{minipage}
\hfill
\end{figure}

Fig.~{\ref{multimodal}} shows an additional instance of the calculations, i.e. energy histograms for $K=26$ out of the $N=64$ for $\alpha = \lambda_{max}$, demonstrating the multimodal distribution which is the result of the interplay between GD-dynamics and discrete projection.

\begin{figure}[H]
\begin{minipage}[c]{0.95\linewidth}
\includegraphics[width=\linewidth]{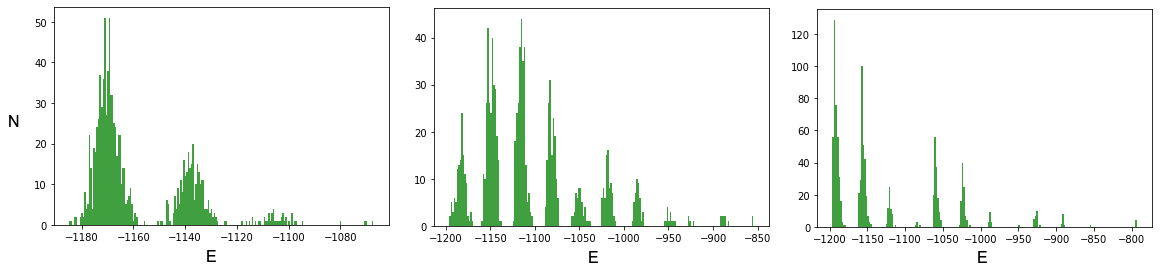}
\caption{Final energy distribution obtained by the numerical integration of Eq.~(\ref{type1}) with $K=28$ out of $N=64$ planted patterns for various $\alpha$ with the final projection on binary states. Left: $\alpha=\lambda_{max}/2$. Middle: $\alpha=\lambda_{max}$. Right: $\alpha=4 \lambda_{max}$. }
\label{K28_N64_projections}
\end{minipage}
\hfill
\end{figure}

\begin{figure}[H]
\begin{minipage}[c]{0.95\linewidth}
\includegraphics[width=\linewidth]{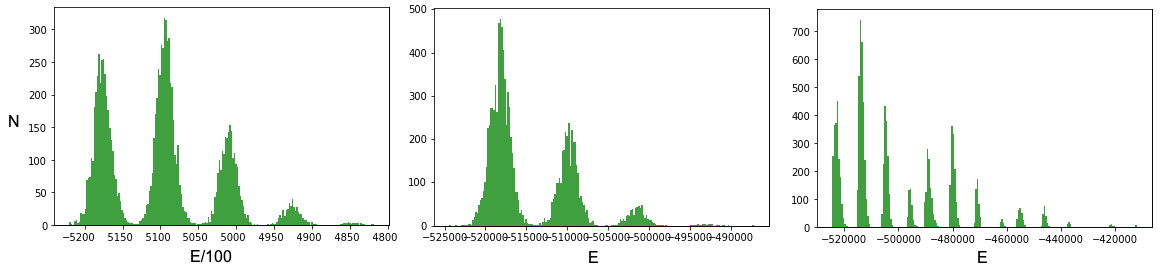}
\caption{Final energy distribution obtained by the numerical integration of Eq.~(\ref{type1}) with $K=28$ out of $N=1024$ planted patterns for various $\alpha$ with the final projection on binary states.. Left: $\alpha=\lambda_{max}/2$. Middle: $\alpha=0.8 \lambda_{max}$. Right: $\alpha=\lambda_{max}$.}
\label{K28_N1024_projections}
\end{minipage}
\hfill
\end{figure}

\begin{figure}[H]
\begin{minipage}[c]{0.95\linewidth}
\includegraphics[width=\linewidth]{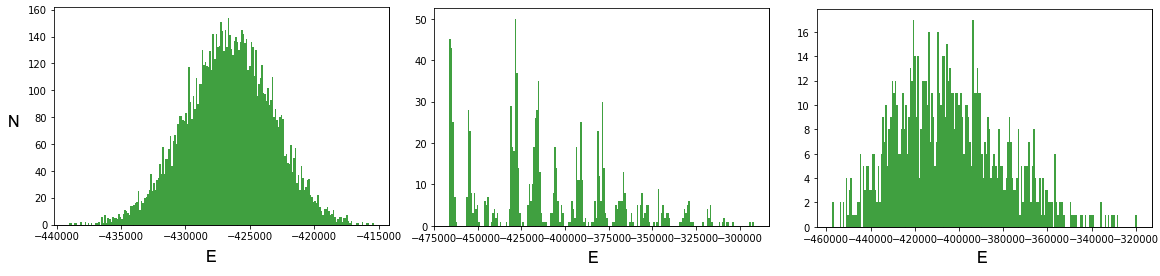}
\caption{Final energy distribution obtained by the numerical integration of Eq.~(\ref{type1}) with $K=500$ out of $N=1024$ planted patterns for various $\alpha$ with the final projection on binary states. Left: $\alpha=\lambda_{max}/2$. Middle: $\alpha= \lambda_{max}$. Right: $\alpha=1.2\lambda_{max}$.}
\label{K500_N1024_projections}
\end{minipage}
\hfill
\end{figure}

\begin{figure}[H]
\begin{minipage}[c]{0.95\linewidth}
\includegraphics[width=\linewidth]{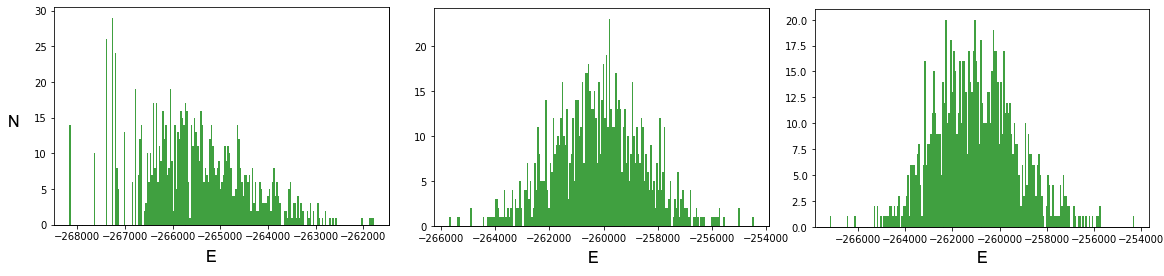}
\caption{Final energy distribution obtained by the numerical integration of Eq.~(\ref{type1}) with $K=1024$ out of $N=1024$ planted patterns for various $\alpha$ with the final projection on binary states. Left: $\alpha=\lambda_{max}/2$. Middle: $\alpha=0.8 \lambda_{max}$. Right: $\alpha=1.5\lambda_{max}$.}
\label{K1024_N1024_projections}
\end{minipage}
\hfill
\end{figure}

Additional plots in Figs.~{\ref{K28_N64_projections}},~{\ref{K28_N1024_projections}},~{\ref{K500_N1024_projections}},~{\ref{K1024_N1024_projections}} present the numerical simulation results to show the influence of both the projection strength $\alpha$ and number of the planted patterns $K$ on the form of the obtained distributions.

\begin{figure}[H]
\begin{minipage}[c]{0.95\linewidth}
\includegraphics[width=\linewidth]{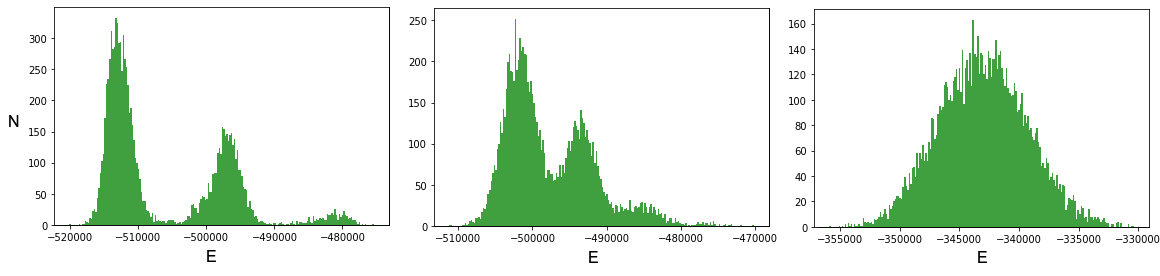}
\caption{Final energy distribution obtained by the numerical integration of Eq.~(\ref{type1}) with $\alpha=\lambda_{max}/2$ and $N=1024$ for a various number of the planted patterns $K$ with the final projection on binary states. Left: $K=50$. Middle: $K=100$. Right: $K=800$.}
\label{lambda_N1024_projections}
\end{minipage}
\hfill
\end{figure}


\subsection{Additional calculations using sampling methods}
\label{Additional calculations using sampling methods}

Here we present additional calculations utilizing sampling methods to illustrate several points. First is to prove that our methodology produces more complex problems than random matrix ensembles (by random problems we mean precisely the comparison with the Ising problem with symmetric coupling coefficients drawn from the normal distribution $\mathcal{N}(0,\,1)$ (or equivalently GUE ensemble). Although this is not the best benchmark in terms of growing complexity, it (complexity) still scales exponentially. To show the difference, we performed numerical simulations using various algorithms (Simulated Annealing method (SA) and Parallel Tempering Monte Carlo (PT)) on both sets, i.e. matrices produced by our main methodology and the GUE(N) ensembles.

\begin{figure}[H]
  \begin{minipage}[c]{0.5\textwidth}
    \includegraphics[width=\linewidth]{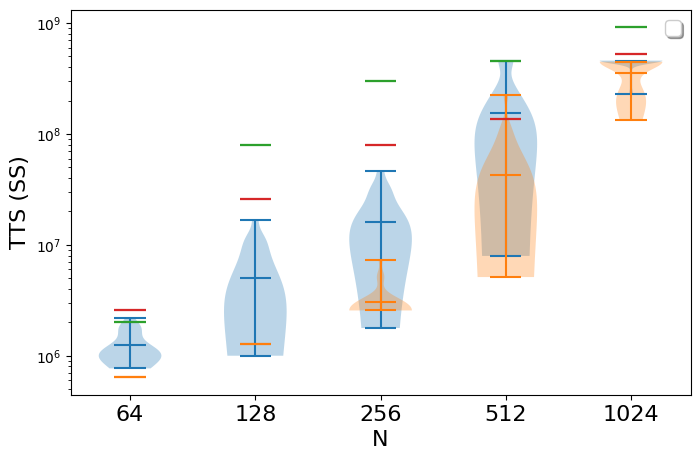}
  \end{minipage}\hfill
  \begin{minipage}[c]{0.4\textwidth}
    \caption{Time-to-solution (in Spin Sweeps) for the set of random symmetric matrices $J_{ij}$. Blue and orange are TTS distributions of the SA and PT algorithms correspondingly. Red and green marks are TTS of the PT and SA algorithms correspondingly for the matrices with the $K=10$ planted patterns. $N=1024$ case is the lower estimate of the TTS.
       } \label{TTS_random}
  \end{minipage}
\end{figure}

Fig.~{\ref{TTS_random}} shows higher time-to-solution (TTS) for our methodology, than for the random matrices. We believe that picking other random baselines requires additional independent work to be carried out.
The details of the corresponding simulations with the random matrices are presented below.

We adjusted the values of the randomly generated matrices by dividing each value by the $\sqrt{N}$, where $N$ represents the size of the matrix. We performed this normalization process for all $20$ matrices generated. For the simulated annealing (SA, blue color), we conducted $ N_{\text{iter}}= 1000000$ spin updates per run. We set the maximum temperature as $T_{\text{max}} = 2$, and $T_{\text{min}} = 0.001 \cdot T_{\text{max}}$. The temperature is updated at each iteration using the multiplier $0.001^{1/N_{\text{iter}}}$. Individual spin updates adhere to the detailed balance condition; if the value $\text{exp}(-\Delta E/T)$ exceeds a randomly generated number, the change is accepted. 

For parallel tempering (PT, orange color), we employed $20$ replicas uniformly distributed in the temperature space starting with the $T_{\text{min}} = 0.1$ and ending at the $T_{\text{max}} = 2$. Neighboring replicas $\{E_i, T_i\}$ and $\{E_{i+1}, T_{i+1}\}$ can be swapped with a probability determined by $\text{min} \{1, \text{exp} [(E_{i+1}-E_i)(1 / T_{i+1}-1 / T_i)]\}$ after $1$ Monte Carlo Sweep, which equals $N$ spin updates. In total, we performed $1000$ Monte Carlo sweeps (MCS). While not thoroughly optimized, these parameters suffice to demonstrate the expected performance of SA and PT on our problem instances. 

TTS is calculated according to the well-known formula:
\begin{equation}
\text{TTS}= \tau \frac{\ln (1-0.99)}{\ln \left(1-p_{\text{gs}}\right)},
\label{TTS}
\end{equation}
where $\tau$ is the average running time of the algorithm and $p_{\text{gs}}$ is the probability of finding the ground state. We measure TTS in SS (spin sweeps) or MCS (Monte Carlo sweeps) $= N \times \text{SS}$. Final TTS is averaged over $20$ matrix instances.

For the matrices with the planted patterns, the parameters are following:
for the simulated annealing (SA, green), we conducted $N_{\text{iter}}= 2000000$ spin updates per run. We set the maximum temperature as $T_{\text{max}} = |E_{\text{min}}|/10$ for $N=64, 128$ and $|E_{\text{min}}|/100$ for $N=256, 512$ and $1024$, and $T_{\text{min}} = 0.025 \cdot T_{\text{min}}$ for $N=64, 128, 256$ and $0.005 \cdot T_{\text{min}}$ for $N=512, 1024$. The temperature is updated at each iteration using the multiplier $0.025^{1/N_{\text{iter}}}$ or $0.005^{1/N_{\text{iter}}}$. Individual spin updates adhere to the detailed balance condition; if the value $\text{exp}(-\Delta E/T)$ exceeds a randomly generated number, the change is accepted. 

For Parallel Tempering (PT, red), we employed $20$ replicas uniformly distributed in the temperature space between $T_{\text{max}}$ and $T_{\text{min}}$. Neighboring replicas $\{E_i, T_i\}$ and $\{E_{i+1}, T_{i+1}\}$ can be swapped with a probability determined by $\min \{1, \exp [(E_{i+1}-E_i)(1 / T_{i+1}-1 / T_i)]\}$ after $1$ Monte Carlo Sweep, which equals $N$ spin updates. In total, we performed $4000$ Monte Carlo sweeps (MCS). Here $T_{\text{max}} = |E_{\text{min}}|/10$ for $N=64$, $128$ and $|E_{\text{min}}|/100$ for $N=256$, $512$, and $1024$, and $T_{\text{min}}$ is $0.025 \cdot T_{\text{max}}$ for $N=64$, $128$, $256$ and $0.005 \cdot T_{\text{max}}$ for $N=512$, $1024$.

\begin{figure} [H]
\centering
\begin{tabular}{cc}
\includegraphics[width=0.45\textwidth]{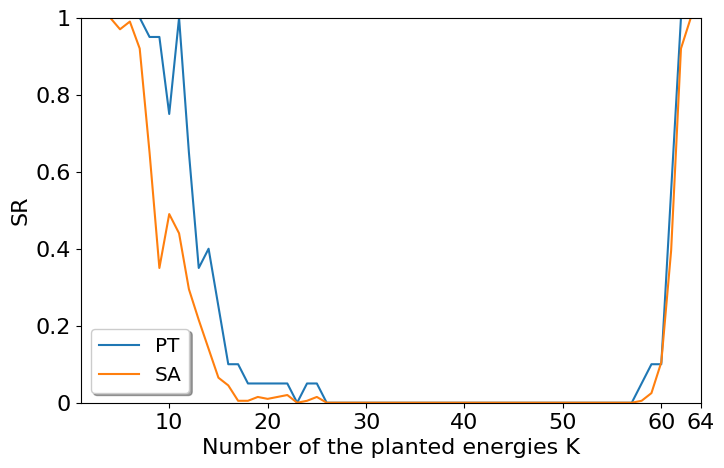} &
\includegraphics[width=0.45\textwidth]{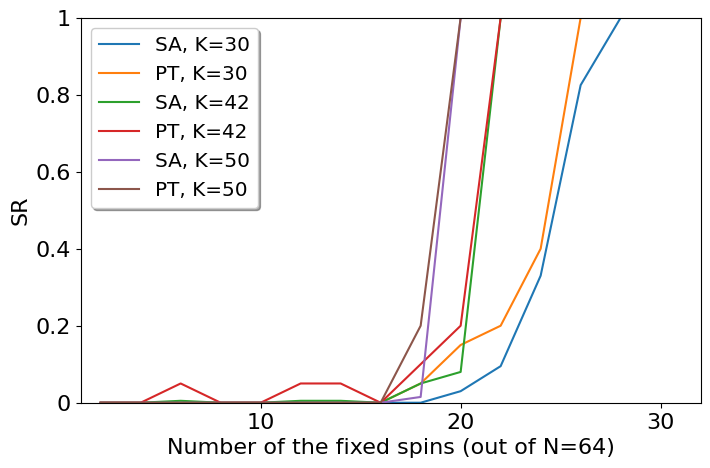} 
\end{tabular}
\caption{Left: Success Rate (SR) of the SA and PT algorithm in the “easy” regime (outside of the $25<K<60$ range). Right: Success rate of the SA and PT algorithm in the “hard” regime ($25<K<60$) on the simplified task. To simplify the search for the ground state, we hold some spins fixed with the values from the known ground state. The x-axis represents the number of fixed spins.}
\label{SR_SA_PT_N64}
\end{figure}

Second point is to show universality of our benchmarks complexity. While our primary focus has been on gradient-based physical systems (and their simulations), the topic of sampling algorithms becomes relevant if one wants to consider sampling machines or hardware, prompting an assessment of their performance on matrices generated by our method. If one doesn't leverage the knowledge that the ground state corresponds to the eigenvector with the highest eigenvalue, a similar complexity pattern emerges. Both algorithms (SA and PT) face challenges in a region akin to the difficulties encountered with Hopfield equations, see Fig.~{\ref{SR_SA_PT_N64}}.

For the Simulated Annealing (SA), we conducted $N_{\text{ITER}} = 2,000,000$ spin updates per run. $T_{\text{max}} = |E_{\text{min}}|/10$, and $T_{\text{min}}=0.025 \cdot T_{\text{max}}$. The temperature is updated at each iteration using the multiplier $0.025^{1/N_{\text{ITER}}}$. Individual spin updates adhere to the detailed balance condition. 

For Parallel Tempering (PT), we employed $20$ replicas uniformly distributed in the temperature space between $T_{\text{max}}$ and $T_{\text{min}}$. Neighboring replicas $\{E_i, T_i\}$ and $\{E_{i+1}, T_{i+1}\}$ can be swapped with a probability determined by $\min \{1, \exp [(E_{i+1}-E_i)(1 / T_{i+1}-1 / T_i)]\}$ after 1 Monte Carlo Sweep. In total, we performed $2000$ Monte Carlo sweeps (MCS). While not thoroughly optimized, these parameters suffice to demonstrate the expected performance of SA and PT on our problem instances. In the easy regime, they exhibit complexity similar to that of the Hopfield equations.

Since obtaining the proper values of both SR and TTS in the "hard" regime is costly, here we solve a simplified version of the original optimization problem. For this purpose, we run the SA and PT algorithms with a certain fixed number of spins, predetermined from the ground state argument. The number of fixed spins serves as a measure along the x-axis, see right part of the Fig.~{\ref{SR_SA_PT_N64}}. The hardest regime is shifted towards higher values of $N$, demonstrating asymmetry. The fixed number of spins contributes an additional term to the optimization assignment, given by the $\sum_i h_i s_i$, with the effective field derived from the fixed spin values represented by the $\sum_j J_{ij} s_j$. For both SA and PT, we utilized the same parameters as in left part of the figure.

\subsection{Decoupling equations}
\label{Decoupling equations}

Using planted patterns with asymmetry coefficients can offer further insights into solver dynamics. Here, we illustrate the primary derivation to support our argument and showcase this additional property of the coupling coefficients.

The main concept revolves around selecting a parametrization for the coupling matrices $J_{ij}$ by representing them as the sum of outer products of specific orthogonal vectors. This approach simplifies the dynamical evolution equations to a point where they are analytically tractable. In other words, it allows us to decouple the matrix-vector multiplication in the linear regime of Equation (\ref{type1}).

For our following derivations, we will use the following notation:
\begin{equation}
x_i=\sum_k a^k \xi_i^k
\label{vector_representation}
\end{equation}
is an example of a particular vector decomposition in the basis $x_i^{k}$ with the corresponding coefficients $a^k $.
The matrix that we choose to work with should have the main property, i.e. to have the following form:
\begin{equation}
J_{ij}=\sum_m^K \lambda^m \xi_i^m \xi_j^m=\sum_m^K \lambda^{m} (\xi^m \otimes \xi^m)_{ij}
\label{matrix_representation}
\end{equation}
which is the sum of the outer products with certain weights taken from the basis vectors, where the index $m$ goes over the $K<N$ planted patterns. One can notice that the coefficients $(w_0+\delta w_m)$ from the Equation (\ref{modified_hebb}) and $\lambda^m$ are the same. The orthogonality of the basis binary vectors $\xi_i^k$ is essential for the decoupling of the equations.

Let us demonstrate the formal derivation of the main equations while making a transition to the new coordinate system in the Eq.~(\ref{type1}) and omitting the nonlinearity ($\varphi(x_j) \approx x_j + O(x_j^2)$):
\begin{equation}
\sum_k \dot a^k \xi_i^k=-\alpha \sum_k a^k \xi_i^k + \sum_j \sum_m^K \lambda^m \xi_i^m \xi_j^m  \sum_k a^k \xi_j^k
\label{initial decomposition}
\end{equation}
The orthogonality property reads (the pattern vectors are not normalised):
\begin{equation}
\sum_j \xi_j^m \xi_j^k=\delta^{m k} N
\label{orthogonality_property}
\end{equation}
It is worth mentioning that the basis vectors for each dimension $N$ are unique up to the permutation between them, for example $(N=4)$:
\begin{equation}
\begin{pmatrix}
\xi^1
 \\\xi^2
 \\\xi^3
 \\\xi^4
\end{pmatrix} = 
\begin{pmatrix}
 1&  1&  1&  1\\
 1&  1&  -1&  -1\\
 1&  -1&  1&  -1\\
 1&  -1&  -1&  1\\
\end{pmatrix} 
\label{basis_vectors_4}
\end{equation}

Summing over $j$ index in Eq.~(\ref{initial decomposition}):
\begin{equation}
\sum_k \dot{a}^k \xi_i^k=-\alpha \sum_k a^k \xi_i^k+\sum_k a^k \sum_m^K \lambda^m \xi_i^m \underbrace{\sum_j \xi_j^m \xi_j^k}_{N \delta^{mk} }
\label{step_1}
\end{equation}
Simplifying the last term:
\begin{equation}
\sum_k a^k \sum_m^K N \delta^{m k} \lambda^m \xi_i^m = 
N \sum_k^K a^k  \lambda^k \xi_i^k
\label{step_2}
\end{equation}
We get:
\begin{equation}
\sum_k^K \dot{a}^k \xi_i^k=-\alpha \sum_k^K a^k \xi_i^k+N \sum_k^K a^k  \lambda^k \xi_i^k
\label{step_3}
\end{equation}

Multiplying on the arbitrary basis vector $\xi_i^ {k^*}$, summing over the $i$ index and dividing by $N$ gives:
\begin{equation}
 \dot a^{k}=-\alpha a^k+N\lambda^k a^k
 \label{core_equation}
\end{equation}
which is the core equation or decoupled mode evolution (for the number of the planted patterns $K$) with the simple exponential function as the solution.

In summary, by selecting the projection of spin variables onto planted patterns and utilizing the existing $J_{ij}$, the Hopfield equations can be effectively decoupled (in the linear regime) into $N$ independent differential equations. These equations can be precisely solved with a few assumptions (by introducing the cut-off time and neglecting or approximating the nonlinearity). The resulting set of independent equations offers a means to describe individual trajectories comprehensively, which allows one to describe the probability flow of the various initial distributions. We believe that current framework allows one to touch many other subjects as well.

\subsection{Planted patterns matrix generation}
\label{Planted patterns matrix generation}

To avoid the confusion and make the description of the generating matrices according to the Eq.~(\ref{modified_hebb}) more clear, we attach Python code that we used to generate matrices with the planted patterns (parameters correspond to the $N=64, K=50, \delta w = 0.001$ case:

To ensure clarity and remove any ambiguity regarding the description of the matrices generation as outlined in the Eq.~(\ref{modified_hebb}), we are providing accompanying \texttt{Python} code. We used this code to generate matrices with the planted patterns (parameters correspond to the $N=64, K=50, \delta w = 0.001$ case):

\begin{lstlisting}[language=Python, caption=Planted patterns matrix generation]
import numpy as np

def create_composite_matrix(binary_matrix):
    rows, cols = np.shape(binary_matrix)[0], np.shape(binary_matrix)[1]
    composite_matrix = np.zeros((2 * rows,2 * cols))
    for i in range(rows):
        for j in range(cols):
            composite_matrix[i][2 * j] = binary_matrix[i][j]
            composite_matrix[i][2 * j + 1] = binary_matrix[i][j]
            composite_matrix[i + rows][2 * j] = binary_matrix[i][j]
            composite_matrix[i + rows][2 * j + 1] = - 1. * binary_matrix[i][j]
    return composite_matrix

degree = 6
N = 2**degree
composite_matrix = np.array([[1,1],[1,-1]])

for j in range(deg-1):
    composite_matrix = create_composite_matrix(composite_matrix)

K = 50
J = np.zeros((N,N)) 
delta_w = 0.001 

for i in range(K):
    J = J - (1.0+delta_w*(i+1))*np.outer(composite_matrix[i],composite_matrix[i]) 
J = J - np.eye(N)*J

print(J)
\end{lstlisting}

\clearpage
\acknowledgements
N.G.B acknowledges the support from the Julian Schwinger Foundation JSF-19-02-0005 and Horizon Europe 10086022. N.D. acknowledges the support from the Israel Science Foundation grant No. 1520/22 and ISF-NSFC grant 3652/21.

\bibliography{references}{}
\bibliographystyle{ieeetr}

\end{document}